\documentclass[12pt,preprint]{aastex}

\usepackage{color}

\def\vec#1{\mbox{\boldmath $#1$}}

\slugcomment{Not to appear in Nonlearned J., 45.}

\shorttitle{Buildup and Release of Magnetic Twist  
           in the AR10930}
\shortauthors{Inoue et al.}

\begin{document}


\title{Buildup and Release of Magnetic Twist during the X3.4 
       Solar Flare of December 13, 2006}
       

\author{S.\ Inoue,}
\affil{School of Space Research,Kyung  Hee University, Yongin, Gyeonggi-do 
       446-701,\\ Republic of Korea}
\email{inosato@khu.ac.kr}

\author{D.\ Shiota,}
\affil{Advanced Science Institute, RIKEN(Institute of Physics and Chemical
       Research), \\  Wako, Saitama 351-0198, Japan}

\author{T.\ T.\ Yamamoto,}
\affil{Solar-Terrestrial Environment Laboratory, Nagoya University \\
       Furo-cho, Chikusa-ku, Nagoya, 464-8601,Japan}

\author{V.\ S.\ Pandey,}
\affil{School of Space Research,Kyung  Hee University, Yongin, Gyeonggi-do 
       446-701, \\ Republic of Korea}

\author{T.\ Magara,}
\affil{School of Space Research,Kyung  Hee University, Yongin, Gyeonggi-do 
       446-701, \\ Republic of Korea}

\author{G.\ S.\  Choe}
\affil{School of Space Research, Kyung  Hee University, Yongin, Gyeonggi-do 
       446-701, \\ Republic of Korea}




  \begin{abstract}
   We analyze the temporal evolution of the three-dimensional (3D) 
   magnetic structure of the flaring active region (AR) NOAA 10930 by 
   using the nonlinear force-free fields extrapolated from the photospheric 
   vector magnetic fields observed by the Solar Optical Telescope on board 
   {\it Hinode}. This AR consisted mainly of two types of twisted magnetic 
   field lines: One has a strong negative (left-handed) twist due to the 
   counterclockwise motion of the positive sunspot and is rooted in the 
   regions of both polarities in the sunspot at a considerable distance 
   from the polarity inversion line (PIL). In the flare phase, dramatic 
   magnetic reconnection occurs in those negatively twisted lines in which 
   the absolute value of the twist is greater than a half-turn. The other 
   type consists of both positively and negatively twisted field lines 
   formed relatively close to the PIL between two sunspots. A strong CaII 
   image began to brighten in this region of mixed polarity, in which the 
   positively twisted field lines were found to be injected within one day 
   across the pre-existing negatively twisted region, along which strong 
   currents were embedded. Consequently, the central region near the PIL 
   distributed with a mix of  differently twisted field lines and the 
   strong currents may play a prominent role in flare 
   onset.
   \end{abstract}

  \section{Introduction}
   Solar flares and coronal mass ejections (CMEs) are eruptive liberation of 
   the accumulated free magnetic energy in the solar corona and are considered 
   the biggest explosions in our heliosphere. These phenomena affect geospace 
   in the form of electromagnetic disturbances called geomagnetic storms. 
   Therefore, it is an important issue for the space weather forecast to have 
   the better understanding of the physical reason responsible for triggering 
   these phenomena.
   
   Many models based on the magnetohydrodynamic (MHD) approach have 
   been proposed and have indicated that twist and shear of the magnetic 
   field lines in particular are responsible for these processes. The twist 
   number of the magnetic field is crucial for analyzing the stability of the 
   magnetic flux tubes. For instance, in a periodic system such as a cylinder,
   a twist with more than one turn could destabilize a flux tube; this is 
   widely known as the Kruskal-Shafranov limit 
   (\citealt{1958PhFl....1..265K}). The dynamics of cylindrical flux tubes in 
   the solar corona have been investigated by \cite{1990JGR....9511919F} and 
   \cite{1993ApJ...417..368I} in two-dimensional (2D) space, and their work 
   was later extended to three-dimensional (3D) space by 
   \cite{2006ApJ...645..742I}. In an anchored flux tube, such as a coronal
   loop, a stronger twist is required to destabilize the ideal MHD modes 
   (\citealt{2004A&A...413L..27T}). Therefore, an accurate quantification of 
   the magnetic twist is necessary in order to clarify the role of the ideal 
   MHD instabilities in triggering solar flares. However, there is still no 
   consensus on the question of how strong the twist and shear of the magnetic 
   field lines must be to trigger a flare.  
   
   Active region (AR) NOAA 10930 produced an X3.4-class flare at 02:10 UT on 
   December 13, 2006, and also generated a coronal mass ejection (CME) that 
   caused electromagnetic disturbances in geospace 
   (\citealt{2008ApJ...689..563L}; \citealt{2009JGRA..11410102K}). 
   Flare-associated features (e.g., the flare ribbon, X-ray sigmoid, and cusp 
   loop structure) in this region were observed well by {\it Hinode} 
   (\citealt{2007SoPh..243....3K}). Furthermore, {\it Hinode} successfully 
   conducted continuous observations of the photospheric magnetic field 
   corresponding to this AR at an unprecedentedly high resolution before and 
   after the X3.4-class flare occurred. Therefore, this is an ideal object to 
   study in order to understand the characteristics of ARs producing X-class 
   flares.  

   The positive sunspot in AR 10930 was associated with strong sheared 
   and twisted counterclockwise motion, whereas the negative sunspot was 
   almost stationary compared to the positive one. As a result of the 
   counterclockwise motion of the positive sunspot, a negative (left-handed) 
   twist was injected into the overlying coronal magnetic loops. 
   \cite{2007PASJ...59S.785S} reported a strong shear field on the basis of 
   X-ray observations in the solar corona. The development of the magnetic 
   energy and its injection into the corona in the form of magnetic helicity 
   because of the strong shearing and twisting motions of the sunspot were 
   also reported recently by
   \cite{2008PASJ...60.1181M},
   \cite{2009ApJ...697L.103S}, and
   \cite{2010ApJ...720.1102P}.

   The apparent motions of the positive sunspot associated with this region 
   were also reported in some previous studies; {\it e.g.},  
   \cite{2007ApJ...662L..35Z} investigated the sunspot rotation using 
   white-light images from the {\it Transition Region And Coronal Explorer} 
   and estimated its rotation for three days beginning on December 11, 2006, 
   finding that it was about $240^{\circ}$. On the other hand, 
   \cite{2009SoPh..258..203M} estimated that it was about $540^{\circ}$ by 
   using G-band images taken by the Solar Optical Telescope (SOT) on board 
   {\it Hinode} (\citealt{2008SoPh..249..167T}). Furthermore, the local area 
   surrounding the polarity inversion line (PIL) contains small-scale complex 
   magnetic structures of mixed polarities due to the twisting motion of the 
   positive sunspot. Some authors also suggested the formation of such 
   structures because of flux emergence around the PIL
   (
   \citealt{2007ApJ...662L..35Z};
   \citealt{2007PASJ...59S.779K};
   \citealt{2008ApJ...687..658W};
   \citealt{2010ApJ...719..403L}
    )
   and indicated that it plays a key role in triggering flares
   (
   \citealt{2010ApJ...720.1102P};
   \citealt{2011ApJ...740...19R}). 
   It is obvious from the above information that the apparent rotation of 
   the sunspot might be rather ambiguous for quantitative measurement of the 
   magnetic field twist. Therefore, it is difficult to obtain appropriate 
   information on the field line twist in 3D space.

   Consequently, nonlinear force-free field (NLFFF) extrapolation becomes 
   a powerful tool for understanding the 3D magnetic structure. 
   \cite{2008ApJ...675.1637S} applied NLFFF extrapolation to AR 10930 
   and identified a strong electric current region above the neutral line 
   before the flare, most of which disappeared as the flare proceeded. 
   \cite{2008ASPC..397..110I} also indicated the possibility of field line 
   relaxation during this flare. Furthermore, \cite{2008ApJ...679.1629G} 
   suggested that a magnetic dip might be formed in the AR and suggested 
   that it could sustain a filament above the magnetic neutral line. 
   \cite{2008ApJ...676L..81J} analyzed the altitude variation in the magnetic 
   structure in the pre- and post-flare phases and found that the energy 
   release process proceeded in some height range from $\sim 8$ Mm to 
   $\sim 70$ Mm, whereas the non-potentiality of the magnetic field increased 
   after the flare below $\sim 8$ Mm. However, there is still no commonly 
   accepted explanation for the entire flare dynamics ({\it i.e.}, from energy 
   accumulation to relaxation) based on the quantitative twist or topology of 
   the field lines.

   \par Recently, \cite{2011ApJ...738..161I} also developed an NLFFF 
    extrapolation procedure based on the MHD relaxation method and applied it 
   to AR 10930. They confirmed its reliability by comparing the location of 
   the footpoints of the sheared field lines across the PIL with that of the  
   CaII illumination obtained by SOT/{\it Hinode}. Their results show that 
   the footpoints before the flare correspond well to the location of the 
   CaII illumination in the central area of the entire domain. They also 
   introduced the magnetic twist, which represents the degree of twist in 
   the magnetic field lines, and clarified that the many strongly twisted 
   lines in this AR have less than a one-turn twist, which indicates 
   robustness against kink instability. The fraction of the magnetic flux 
   corresponding to the strongly twisted lines having more than a one-turn 
   twist was found to be negligibly small compared to those of less twisted 
   ($\sim $0.5- to $\sim$ 1.0-turn twist) field lines. On the other hand, in 
   a later study, \cite{2012ApJ...747..65I} investigated the 3D magnetic 
   structure making up the sigmoid and clarified the relationship between the 
   X-ray intensity obtained from the X-Ray Telescope (XRT) on {\it Hinode} 
   and the magnetic twist or field-aligned current obtained from the NLFFF. 
   They found that strong X-ray intensity is closely related to the 
   field-aligned current flowing in the chromosphere rather than the twist 
   values of the strongly twisted field lines. They further indicated that 
   the field line patterns generated by the NLFFF are quite similar to the 
   profiles obtained in the flux-emergence simulation of 
   \cite{2004ApJ...605..480M}. In addition, other topics, such as the 
   characteristics of the 3D magnetic field in this AR, the formation process 
   of flare-producing ARs, and a quantitative interpretation of the flare 
   dynamics in terms of the magnetic twist, were not addressed in our 
   previous studies.
     
   In this paper, we also analyze the magnetic twist of AR 10930 by applying 
   NLFFF extrapolation to time series vector magnetogram data obtained 
   from {\it Hinode} in order to understand these problems. The rest of the 
   paper is structured as follows. The numerical model and data set are 
   described in Section 2. The results of our analysis of the 3D structure, 
   magnetic twist, and topology of AR 10930 are presented in Section 3. 
   Their implications for the flare onset mechanism are discussed in 
   Section 4, and the key conclusions are summarized in Section 5.

  \section{Numerical Method}
  \subsection{Extrapolation of the NLFFF}
   We adopt the same force-free extrapolation method as in our previous 
   studies ({\it e.g.}, \citealt{2011ApJ...738..161I} and 
   \citealt{2012ApJ...747..65I}), which employed the MHD relaxation method.   
   A 3D field ideally has to be extrapolated so that it satisfies a 
   force-free condition, such as
   \begin{equation}
   \vec{\nabla} \times \vec{B} = \alpha \vec{B}. 
   \label{eq_ff}
   \end{equation}
   Unfortunately, the photospheric field obtained from an observation is 
   not necessary to satisfy the force-free condition. Nevertheless, 
   \cite{2011ApJ...738..161I} and \cite{2012ApJ...747..65I} carefully 
   evaluated the reliability of an extrapolated field by comparing it with 
   CaII and X-ray images from the SOT and XRT on board {\it Hinode}. They 
   found that their results are quite consistent with the observations. 
   \cite{2009ApJ...691..105S}, 
   \cite{2009ApJ...703.1766S}, and 
   \cite{2011JGRA..11601101H} 
   also compared their force-free field with multi-wave observations. The 
  magnetic field lines obtained from their results were also found to be 
  quite reasonable for recapturing the X-ray or EUV images obtained by 
  {\it Hinode}; hence, these results support the suggestion that the NLFFF 
  is a solid tool for describing 3D magnetic structure. 
  
  In this study, the solved equations are governed by the MHD-like equations
  for a low-$\beta$ plasma,

  \begin{equation}
  \frac{\partial \vec{v}}{\partial t} 
                        = - (\vec{v}\cdot\vec{\nabla})\vec{v}
                          + \frac{1}{\rho} \vec{J}\times\vec{B}
                          + \nu\vec{\nabla}^{2}\vec{v},
  \label{motion}  \end{equation}
  \begin{equation}
  \frac{\partial \vec{B}}{\partial t} 
                        =  \vec{\nabla}\times(\vec{v}\times\vec{B}
                          -\eta\vec{J})
                          -\vec{\nabla}\phi, 
  \label{induc_eq}
  \end{equation}

  \begin{equation}
  \vec{J} = \vec{\nabla}\times\vec{B},
  \end{equation}

    \begin{equation}
    \frac{\partial \phi}{\partial t} + c^2_{h}\vec{\nabla}\cdot\vec{B} 
= -\frac{c^2_{h}}{c^2_{p}}\phi,\label{div_eq1}
   \end{equation}
   where the last equation was originally introduced by
  \cite{2002JCoPh.175..645D} for calculating an MHD solution that 
  also avoids deviation from the solenoidal condition 
  $\vec{\nabla}\cdot\vec{B} =0$. Here, $\vec{B}$ is the magnetic flux 
  density, $\vec{v}$ is the velocity, $\vec{J}$ is the electric current 
  density, $\rho$ is the pseudo-density (which is assumed to be 
  proportional to $|\vec{B}|$), and $\phi$ is the convenient potential. 
  The length, magnetic field, velocity, time, and electric current density 
  are normalized by 
  $L_0=5.325 \times 10^{9}$ (cm), 
  $B_0=3957$ (G), 
  $V_{A}\equiv B_{0}/(\mu_{0}\rho_{0})^{1/2}$, 
  $\tau_{A}\equiv L_{0}/V_{A}$, 
  and 
  $J_0=B_{0}/\mu_{0} L_{0}$, 
  respectively. The non-dimensional viscosity $\nu$ is fixed at 
  $(1.0\times 10^{-3})$, and the non-dimensional resistivity $\eta$ 
  is assumed to have the functional form given in \cite{2011ApJ...738..161I} 
  and \cite{2012ApJ...747..65I}, 
  \begin{equation}
  \eta = \eta_0 + \eta_1 \frac{|\vec{J}\times\vec{B}||\vec{v}|^2}{\vec{|B|}},
  \end{equation} 
  where $\eta_0 = 5\times 10^{-5}$, and $\eta_1 = 1.0 \times 10^{-3}$, in 
  non-dimensional units. The other parameters, $c_h^2$ and $c_p^2$, are  
  fixed at 0.04 and 0.1, respectively.

  The magnetogram set on the bottom boundary is a hybrid map from 
  SOT/{\it Hinode} and the Michelson Doppler Imager (MDI) on the 
  {\it Solar  and Heliospheric Observatory} ({\it SOHO}) 
  (\citealt{1995SoPh..162..129S}). To obtain an extensive overview of 
  the numerical box (see Figure \ref{f2}(a) in \citealt{2012ApJ...747..65I}), 
  the magnetogram from the spectropolarimeter (SP) on {\it Hinode} is 
  located at the center of the MDI/{\it SOHO} data. The area outside 
  of the region covered by the SP magnetogram is maintained by the 
  longitudinal field of MDI/{\it SOHO}, whereas its tangential components 
  are fixed by the potential field derived on the basis of a synoptic 
  chart of the line-of-sight component of the magnetic field observed 
  by MDI/{\it SOHO}. The lateral and upper boundary conditions are also 
  fixed by the potential field components.   

  An initial condition is given by the potential field calculated from the 
  normal component of the magnetic field on all the boundaries after 
  satisfying $\int B_n dS = 0$ in the entire domain, where $dS$ 
  represents the surface element on all the boundaries, and the subscript 
  n represents the component normal to the surfaces of the boundaries. 
  The 3D configuration is shown in Figure \ref{f2}(b) of 
  \cite{2012ApJ...747..65I}. The velocity field {\vec{v}} set to zero on 
  all the boundaries. A Neumann-type boundary condition 
  ($\partial_n \phi = 0$) is applied to the potential $\phi$ at all the 
  boundaries, where $\partial_n$ represents the derivative along the normal 
  direction on the surface.

  The numerical scheme for this calculation is given by the Runge-Kutta-Gill 
  scheme with fourth-order accuracy for the temporal integral and the 
  central finite difference with second-order accuracy for the spatial 
  derivative. The simulation domain is a rectangular box spanning 
  $(0, 0, 0)<(x, y, z)< (4.0L_0, 4.4L_0, 2.2L_0)$, which corresponds to 
  $(295.2'', 324.6'', 162.3'')$ in view angle. The domain is uniformly 
  divided by a $128 \times 128 \times 64$ grid. The vector-field 
  magnetogram located on the bottom boundary ($125\times 64$ grid) 
  was formed by binning from the original magnetogram, which had a 
  $1000 \times 512$ grid.

  \subsection{Observations}
  We extrapolated the 3D force-free field of AR 10930 by using the 
  vector magnetograms of an observation provided by the SP on 
  SOT/{\it Hinode}. We used five magnetograms, which were observed 
  at 17:00 UT on December 11; at 03:50 UT, 17:40 UT, and 20:30 UT on 
  December 12; and at 04:30 UT on December 13, 2006. The first four 
  data sets were taken before the onset of the X3.4-class flare, 
  which occurred at 02:10 UT on December 13, 2006, whereas the last 
  one was observed just after it. These data were obtained by 
  Milne-Eddington inversion of the FeI lines at 630.15 nm and 630.25 nm. 
  The minimum energy method was applied to solve the $180^{\circ}$ 
  ambiguity (\citealt{1994SoPh..155..235M}; \citealt{2006SoPh..237..267M};   
  \citealt{2009ASPC..415..365L}). We also compared the CaII H 3968 
  $\AA$ images with the 3D NLFFF. These images were provided by 
  the Broadband Filter Imager (BFI) on SOT/{\it Hinode}, which has a 
  time cadence of 2 min. The field of view is 223.15''$\times$111.58'', 
  with a pixel size of about 0.109''. The CaII line is sensitive to 
  temperature at  $10^4$K, which corresponds to the lower chromosphere, 
  and reacts strongly to chromospheric heating. In this study, the 
  CaII images were employed from just after the flare onset to the growth 
  phase of the two-ribbon structure, which corresponds to the time interval 
  from 02:14 UT to 02:40 UT on December 13.

  \subsection{Analysis of the magnetic twist}
  We focus on the magnetic twist, which is defined as the turnover number 
  of magnetic field lines in magnetic flux tubes and plays a key role in 
  judging the stability or instability of a magnetic configuration, as 
  mentioned in section 1. Although the detailed formulation and its 
  implementation in terms of the results obtained from the NLFFF on AR 
  10930 using this magnetic twist are illustrated in our previous works, 
  ({\it e.g.}, \citealt{2011ApJ...738..161I} and 
  \citealt{2012ApJ...747..65I}), we briefly describe this key issue again 
  here to refresh readers memories. 

  The magnetic twist ($T_n$) is related to the magnetic helicity; {\it e.g.}, 
  the helicity of a closed flux tube anchored on the solar surface is 
  described by the following equation (
  \citealt{1984JFM...147..133B}; 
  \citealt{1992RSPSA.439..411M};
  \citealt{2006JPhA...39.8321B}), 
  \begin{equation} 
    H = (T_n + W_r){\Phi}^2,
  \end{equation} 
  where {\it H}, $\Phi$, and $W_r$ are the magnetic helicity, magnetic 
  flux of a cross section of the flux tube, and magnetic writhe corresponding 
  to the helical structure of an axis of the field line, respectively. $T_n$ 
  indicates how much of the magnetic helicity is generated by the currents 
  parallel to the flux tube axes $J_{||}$; thus, $T_n$ is described by  
  \begin{equation}
   T_n = \int \frac{dT_n}{ds}ds = \int \frac{J_{||}}{4 \pi B_{||}} ds,
  \label{eq_org_twist}
  \end{equation}
  where $||$ indicates the component parallel to the field line, and the line 
  integral $\int ds$ is taken along the central magnetic field line of the 
  flux tube (\citealt{2006JPhA...39.8321B}; \citealt{2010A&A...516A..49T}).
 
   In this study, because we assume the force-free state and calculate 
  $T_n$ for each field line, the twist value can be replaced by 
  \begin{equation}
    T_n = \frac{1}{4\pi} \int \alpha dl  =   \frac{1}{4\pi} \alpha L,
  \label{eq_twist}  
  \end{equation}
  where $\alpha$ is the force-free $\alpha$ given in equation (\ref{eq_ff}), 
  the line integral $\int dl$ is taken along the magnetic field line, and $L$ 
  represents the field line length.

  We introduce the average force-free $\alpha$, which is defined as
  \begin{equation}
  \bar{\alpha} = \frac{1}{2} (\alpha_{+}+\alpha_{-}), 
  \label{eq_ave_alpha}
  \end{equation} 
  where $\alpha_{+}$ and $\alpha_{-}$ indicate the force-free $\alpha$ 
  on the footpoints of each field line on the positive and negative 
  polarities, respectively. This formulation is adopted to minimize the 
  deviation of the force-free condition in the photosphere. Consequently, 
  the twist is formulated in this study as
  \begin{equation}
   T_n = \frac{1}{4\pi} \bar{\alpha} L.
  \end{equation}
  Before calculating the average force-free $\alpha$ ($\bar{\alpha}$), we 
  first focus on those specific magnetic fields for which the normal component 
  exceeds 30 G in order to avoid numerical errors; further, $\alpha = 0$ is 
  assumed for the rest of the area. Next, we average $\alpha$ over 
  $3 \times 3$ cells, and finally we obtain the average force-free 
  $\alpha$ ($\bar{\alpha}$) after calculating $\alpha_{+}$ and $\alpha_{-}$. 

  Note that we should use this formulation of the twist with caution, as 
  given in equations (\ref{eq_org_twist}) or (\ref{eq_twist}). Because its 
  value also depends on the field line length, one would obtain very large 
  values corresponding to the very long field lines that have one footpoint 
  outside the AR. However, in this study we do not consider such long field 
  lines and essentially restrict our analysis to closed field lines, using 
  the original definition in which the footpoints of the magnetic field lines 
  are anchored somewhere inside the AR, and the magnetic twist is considered 
  as being effectively propagated along the field lines through the strong 
  shearing  and twisting motions of the sunspot.

  \section{Results} 
  \subsection{3D structure of the Solar Active Region 10930}
\begin{figure}
  \epsscale{1.}
  \plotone{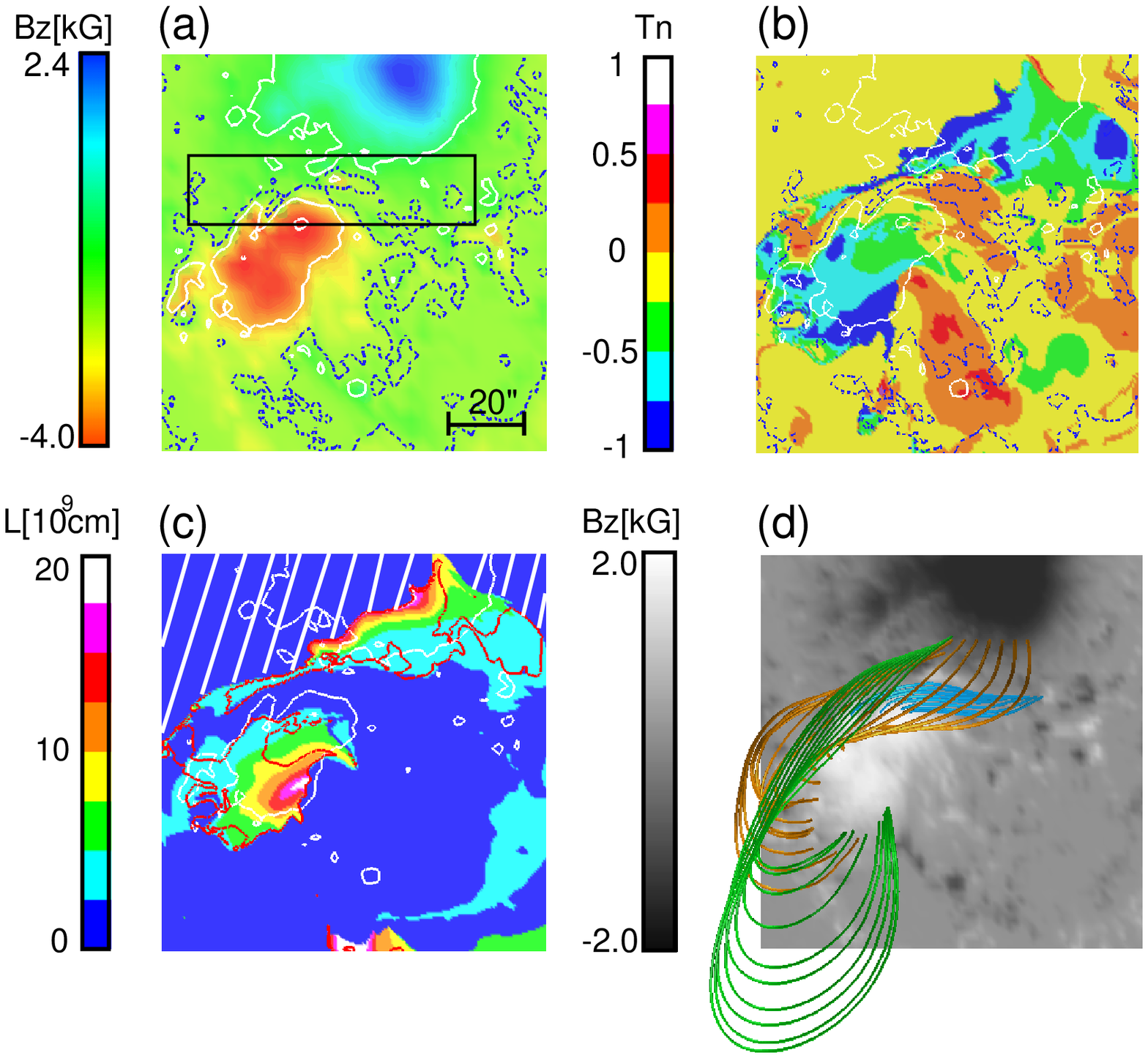}
  \caption{
           (a) Distribution of the normal component of the magnetic field 
               ($B_z$) on the photosphere at 20:30 UT on December 12 
               (6 h before the flare). White solid lines indicate selected 
               contours of $B_z$ (=790 G and $-$790 G); blue dotted lines 
               are the PIL. Black solid square indicates the region near the 
               PIL between the positive and negative polarities.
           (b) Twist ($T_n$) distribution of the field lines from NLFFF at 
               20:30 UT on December 12 mapped on the photosphere with the 
               same white and blue contours as in (a). Positive and negative 
               signs represent the distributions of right- and left-handed 
               twisted lines, respectively.
           (c) Field line length mapped on the photosphere. Thin white lines 
               have the same meaning as in (a); thick white diagonal lines 
               indicate the region occupied by the open field lines that are 
               connected outside the field of view. Red lines indicate a twist 
               magnitude of $|T_n|=0.5$; {\it i.e.}, the regions surrounded 
               by red lines represent $|T_n|>0.5$. 
           (d) Selected field lines plotted over the $B_z$ distribution. Blue, 
               orange, and green field lines correspond to twist ranges of 
               $|T_n|<$0.5, 0.5$< |T_n| <$1.0, and $|T_n|>$1.0, respectively. 
               All panels(a$-$d) have the same field of view.
           }
  \label{f1}
  \end{figure}

   First, we try to understand the characteristics of the 3D magnetic 
   structure 6 h before the flare associated with AR 10930 using twist 
   analysis. The choice of this time is obvious because the vector field 
   in its final pre-flare phase was observed around this time. The details 
   of the force-freeness of the extrapolated fields were given in our 
   previous work, \cite{2012ApJ...747..65I}. Figure \ref{f1}(a) shows the 
   distribution of the normal component of the magnetic field with the 
   selected contours in white solid lines corresponding to $B_z$ = 790 G 
   and $-$790 G, whereas the blue dotted lines correspond to the PIL. 
   The black solid square box indicates the region surrounding the PIL 
   between the positive and negative polarities of the sunspot.
   
   Figure \ref{f1}(b) represents the twist distribution of the field lines 
   of the NLFFF mapped on the photosphere in the same field of view as in 
   Figure \ref{f1}(a), where the white solid and blue dotted lines have the 
   same meaning. Note that these twisted values are focused on closed 
   field lines only, as mentioned in subsection 2.3. The positive and 
   negative twist distributions describe right- and left-handed field line 
   twists, respectively. A strong negative twist ($|T_n|>0.5$) is clearly 
   distributed on both sunspots; the strongest twisted region (more 
   than one turn) also appears here. Hereafter, ``strong twisted lines'' 
   usually refers to field lines having more than a half-turn twist 
   ($|T_n|>0.5$). From this result, it is evident that the energy 
   accumulated field lines are connected mainly between these strong twisted 
   regions, which are at a considerable distance from the PIL.
   On the other hand, the region near the PIL surrounded by the black 
   solid square in Figure \ref{f1}(a) is of mixed polarity, containing both 
   positively and negatively twisted field lines. However, the magnitudes 
   of the twist values are much smaller ($|T_n|<0.25$) than those of the 
   outer field lines characterized by strong negative twist that occupy 
   the main twisted region of both polarities
   
   Figure \ref{f1}(c) shows the field line length ($L$) mapped on the 
   photosphere in the same field of view as Figure \ref{f1}(a), where the 
   regions surrounded by red lines correspond to a twist value of 
   $|T_n| > 0.5$. The region marked by diagonal lines is occupied by open 
   field lines that are connected outside the field of view. The mixed region 
   of positively and negatively twisted field lines near the PIL enclosed by 
   the solid square in Figure \ref{f1}(a) consists of shorter magnetic field 
   lines compared to the strongly twisted region surrounded by the red 
   contours. From these results, AR 10930 can be roughly divided into two 
   regions. One is occupied by the strongly negatively twisted field lines of 
   single magnetic helicity and is located at a considerable distance from 
   the PIL. The other corresponds to a mixed region of both positively and 
   negatively twisted field lines near the PIL between the positive and 
   negative polarities of the sunspot.  
   
    A 3D representation of the magnetic field lines is shown in 
   Figure \ref{f1}(d), which covers the same field of view as 
   Figure \ref{f1}(a). The blue, orange, and green field lines represent 
   selected 3D magnetic field lines having twist values of 
   $|T_n|<0.5$, $0.5<|T_n|<1.0$, and $1.0<|T_n|<1.5$, respectively. The 
   orange and green lines are clearly longer than the blue one. On the 
   other hand, the magnetic shear corresponding to the blue lines is 
   stronger than that of the other field lines despite its weaker twist 
   strength. This will be discussed in more detail in a later section.  

  \subsection{An Evolution of the Magnetic Twist in the Solar Active Region 
              10930}
  \subsubsection{An Evolution of the Strong Twisted field Lines Forming  
                 at a Considerable Distance from the PIL}
\begin{figure}
  \epsscale{.85}
  \plotone{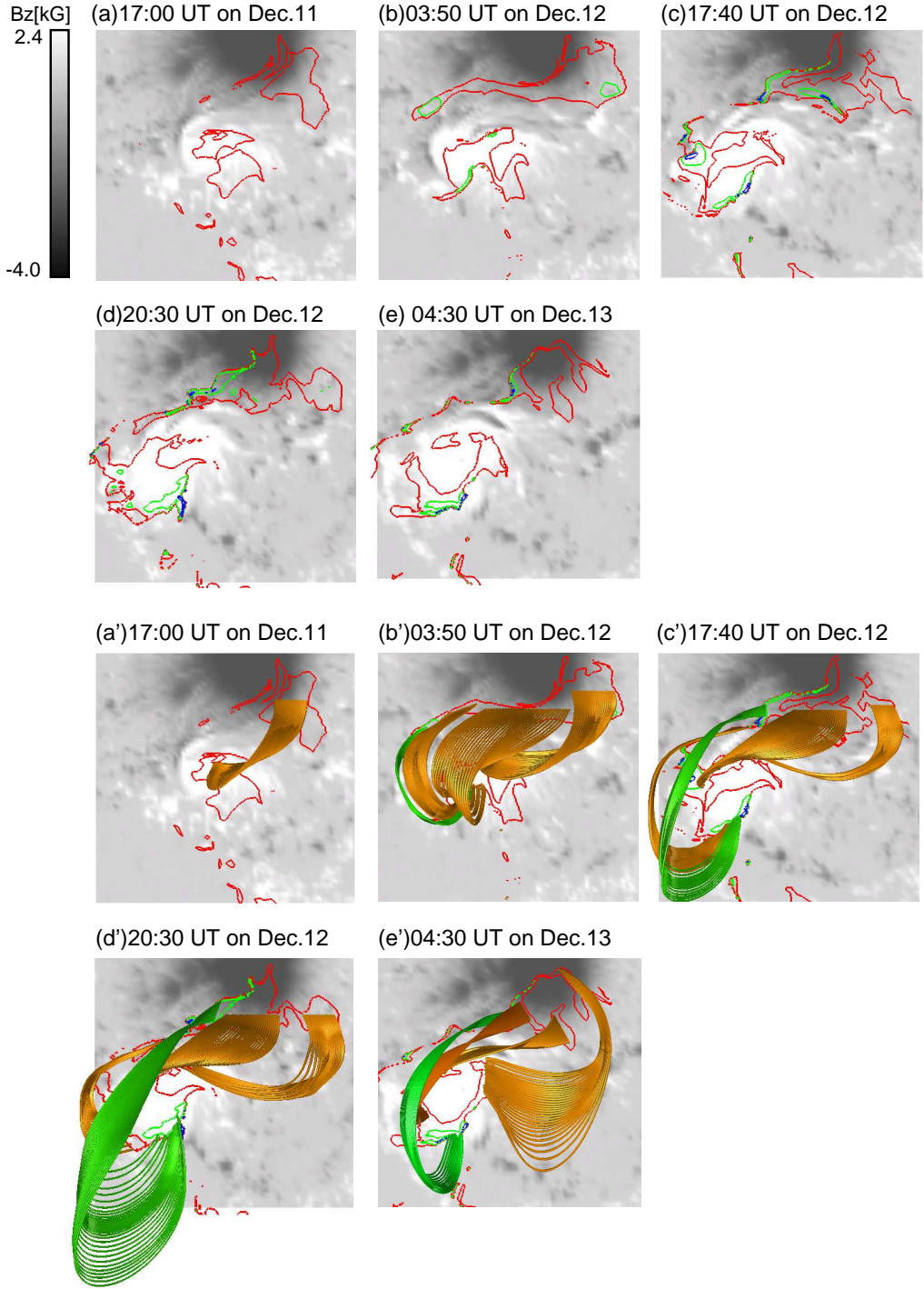}
  \caption
          {
           Upper panels(a$-$e): temporal evolution of selected twist 
           magnitudes $|T_n|=0.5$, $|T_n|=1.0$, and $|T_n|=1.5$, plotted in 
           red, green, and blue lines, respectively, over the normal 
           component of the magnetic field in gray scale in the same field of 
           view as Figure \ref{f1}. 
           Lower panels(a'$-$e'): selected magnetic field lines are also 
           plotted. Orange and green field lines represent twist ranges of 
           $0.5 < |T_n| < 1.0$ and $1.0<|T_n|<1.5$, respectively.           
           }
  \label{f2}
  \end{figure}

  In the previous section, we investigated the 3D magnetic structure 
  of AR NOAA 10930 6 h before the flare. Here, we investigate the temporal 
  evolution of the magnetic twist associated with this AR using the NLFFF 
  derived from the time series of the vector fields taken before and after 
  the flare. In the upper panels of Figure \ref{f2}(a$-$e), the temporal 
  evolution of twist having three different values, $|T_n|=0.5$, $|T_n|=1.0$, 
  and $|T_n|=1.5$, is plotted in red, green, and blue contours, respectively, 
  over the normal component of the magnetic field, represented in gray scale. 
  The field of view is the same as in Figure \ref{f1}. In the lower panels 
  (a'$-$e'), the orange and green lines represent selected 3D magnetic field 
  lines having twist strengths of $0.5 < |T_n| <1.0$ and $1.0 <|T_n| <1.5$, 
  respectively.  

  At 17:00 UT on December 11, the strong twist ($|T_n| > 0.5$) was 
  distributed on the positive sunspot and the southwest edge of the 
  negative sunspot. It was built up by the counterclockwise motion of 
  the positive sunspot, as mentioned in section 1. The strongly twisted 
  regions are connected with orange field lines, as shown in the 
  Figure \ref{f2}(a'). At 03:50 UT on December 12, the twist 
  distribution grew wider, and in some regions, even stronger twist 
  (more than one turn) can be seen. A comparison of the twist analysis 
  at three different times before the flare, 03:50 UT, 17:40 UT, and 
  20:30 UT on December 12, reveals that the regions having a twist of 
  more than one turn were spreading more at both polarities, and 
  some magnetic lines with strong twist appear in the expansion stage 
  owing to the continuous shearing and twisting motions of the positive 
  sunspot. Figure \ref{f2}(d), obtained 6 h before the flare, is the final 
  picture of the magnetic configuration in the pre-flare phase. Although 
  the strongly twisted lines ($|T_n|>1.0$) are widely distributed compared 
  to the previous times, they are localized on the edge of the sunspot, 
  where the magnetic field strength is weaker than at the central part of 
  the sunspot. The post-flare image, Figure \ref{f2}(e), corresponds to 2 h 
  after flare onset. The twisted field lines decrease, which is also reported 
  for another AR in a recent paper \cite{2012ApJ...748...77S}, in the 
  central part of the positive sunspot, which consisted mainly of field 
  lines of twist $|T_n| < 1.0$ before the flare. However, the part of the 
  strong twist confined at the edge of the positive sunspot remained even 
  after the flare.

\begin{figure}
  \epsscale{1.}
  \plotone{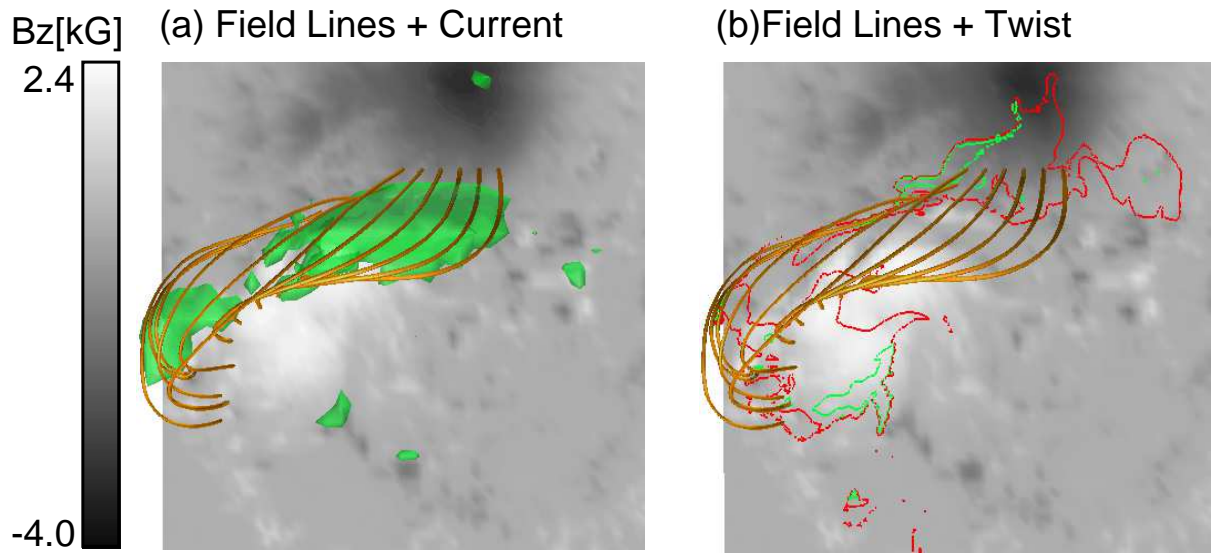}
  \caption
          {
          Magnetic field lines and surface having strong current density 
          ($|J| >$ 2.4) at 20:30 UT on December 12 (orange and green, 
          respectively) over the normal component of the magnetic field 
          (gray scale) in the same field of view as Figure \ref{f1}. 
          (b) Twist distributions in red ($|T_n|=0.5$) and green 
          ($|T_n| = 1.0$) contours with the field lines as in (a).           
          }
  \label{f3}
  \end{figure}

  From these analyses, we infer that although the positive sunspot is 
  capable of generating strongly twisted lines of more than half a turn 
  in the energy accumulation process because of its strong twisting 
  and shearing motions, the strongest twist (in particular, $|T_n|>1.0$) 
  does not seem to contribute efficiently to the main energy release. 
  To further investigate this point, let us examine the distribution of 
  the current density along the twisted field lines. An estimation of the 
  strong current density, which is related to the measurement of a large 
  free energy through the twisted field lines, would provide a way to 
  clarify whether the most strongly twisted lines ($|T_n| > 1.0$) are 
  capable of causing X-class flares. Figure \ref{f3}(a) shows the field 
  lines in orange with the strong current density region represented 
  by the green surface over the normal component of the magnetic 
  field in gray scale at 20:30 UT on December 12. Because these field 
  lines surround the surface of the strong current density region, they 
  are expected to store considerable free energy. Figure \ref{f3}(b) 
  represents the twist distribution by red ($|T_n|=0.5$) and green 
  ($|T_n| = 1.0$) contours mapped on the photosphere, with the same 
  field lines as in Figure \ref{f3}(a). The footpoints of most field 
  lines are clearly rooted in the regions surrounded by the red contours, 
  which clarifies that the twist values of the field lines supporting the 
  strong current density are distributed in the range of $0.5<|T_n|<1.0$. 
  Hence, these results suggest that the field lines having a twist of more 
  than one turn could not be deeply connected with the release of free 
  energy in X-class flares.
    
  \subsubsection{An Evolution of the Twisted Lines Near the Polarity 
   Inversion Lines}
  In this subsection, we present the temporal evolution of the magnetic 
  twist near the PIL. Because many flares are observed to originate at 
  the PIL, it becomes imperative to observe the behavior of the magnetic 
  twist near this region. The left panels in Figure \ref{f4} shows the 
  temporal evolution of the twist distribution near the PIL. Blue and red 
  lines indicate the locations of the negative and positive sunspots, 
  respectively, which are defined as in Figure \ref{f1}. The green lines 
  represent the PIL. The black and white areas represent the negatively 
  and positively twisted regions occupied by closed field lines, whereas 
  the regions occupied by open field lines are shown in gray and are outside 
  the scope of discussion in this study. At 17:00 UT on December 11, this 
  active region consisted almost entirely of field lines having negative 
  twist values. The positive sunspot's motion had already developed this 
  twisted region more than a day before flare onset. The positively twisted 
  regions gradually increased as the flare onset approached. Finally, 
  after the flare, most regions near the PIL between the positive and 
  negative sunspots were occupied by positively twisted field lines. The 
  time evolution of field lines in this AR is characterized mainly by the 
  negatively twisted field due to the positive sunspot's motion 
  (Figure \ref{f2}). However, positive twist begins to buildup at 03:50 UT 
  on December 12, about one day before flare onset, and it formed in less 
  time than the strongly negatively twisted regions. This result is 
  consistent with \cite{2010ApJ...720.1102P}, which describes the injection 
  of positive helicity into a pre-existing system having negative helicity 
  on a shorter time scale. Recently, \cite{2011ApJ...740...19R} also showed 
  the injection of an opposite vertical current near the PIL through the 
  net current distribution on both polarities.     
   
   We also check the average force-free $\alpha$, denoted as $\bar{\alpha}$. 
  The color map in the right panels in Figure \ref{f4} shows the temporal 
  evolution of $\bar{\alpha}$. Its pattern is also very similar to that of 
  the twisted field lines shown in the left panels. The negative 
  $\bar{\alpha}$ is already injected in the early phase, whereas the positive 
  $\bar{\alpha}$ increases gradually near the PIL as time proceeds. Unlike 
  the twist distribution, some of the positive $\bar{\alpha}$, for example, 
  the region surrounded by the solid square in the right panels, is comparable
  in strength to the negative $\bar{\alpha}$, although the magnitudes of the 
  positive twist values are much smaller than those in the negative 
  $\bar{\alpha}$ regions (see Figure \ref{f1}(b)). Consequently, we see a 
  much different picture in terms of the twist analysis.
      
  \subsection{Release Process of the Magnetic Twist in the X3.4 Class Flare 
              Occurred in the Solar Active Region 10930}  

  \subsubsection{Twist distribution vs. Evolution of the Flare Ribbon}
  Here, we investigate the release process of the magnetic twist before 
  and after the flare in detail in order to understand the flare dynamics 
  in terms of the variations in the twist value. First, we check the 
  temporal evolution of the CaII illumination obtained from SOT/{\it Hinode} 
  and compare it with the magnetic twist  obtained through the NLFFF to 
  clarify the relationship between them. 
  
\begin{figure}
  \epsscale{1.}
  \plotone{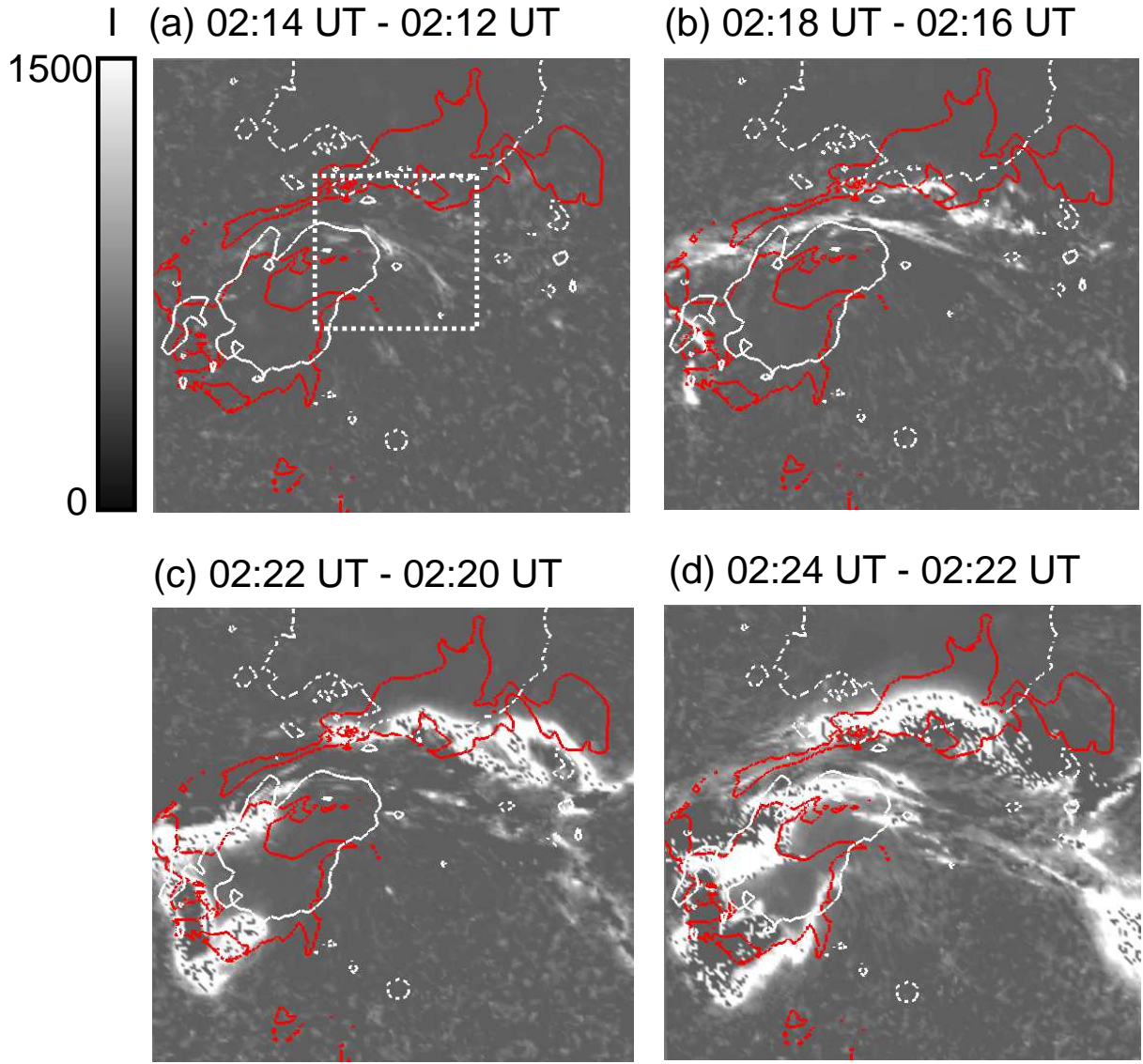}
  \caption{
           (a)-(d) Gray scale maps showing the temporal evolution of 
           CaII images, which are running difference images, 
           corresponding to growth of the flare ribbon in the early phase 
           of the flare (02:14 UT to 02:24 UT on December 13). The field 
           of view is the same as in Figure \ref{f1}. Red lines indicate 
           contours of twist $|T_n|=0.5$ obtained from the NLFFF at 20:30 UT 
           on December 12. White solid and dashed lines represent contours 
           of the normal component of the magnetic field, defined as in 
           Figure \ref{f1}(a). Dotted square suggests the location of 
           illumination corresponding to the flare onset.         
          }
  \label{f5}
  \end{figure}

  The gray scale map in Figure \ref{f5} shows running difference images of 
  CaII from 02:14 UT to 02:24 UT on December 13. This time period corresponds 
  to the growth phase of the two-ribbon structure. The white solid and dashed 
  lines represent the locations of the positive and negative sunspots at 
  20:30 UT on December 12, which are defined as in Figure \ref{f1}. At the 
  initial onset phase, around 02:14 UT on December 13, the CaII image shows 
  strong illumination around the local area nearby the PIL, which is 
  surrounded by the white dotted square. The two-ribbon structure had still 
  not formed at this time. A thin two-ribbon structure started to form around 
  02:18 UT on December 13 and finally became thick and showed strong 
  illumination from 02:22 UT to 02:24 UT on December 13.

  Figure \ref{f5} shows contours corresponding to the twist value $|T_n|=0.5$ 
  (red lines), which were obtained from the NLFFF at 20:30 UT on December 12. 
  The regions surrounded by the red lines are occupied by strongly twisted 
  field lines (greater than a half-turn twist, {\it i.e.}, $|T_n|>0.5$). The 
  initial illumination at 02:14 UT on December 13 clearly comes from the 
  weakly twisted region where $|T_n| < $ 0.25. Although the initial 
  two-ribbon structure at 02:18 UT on December 13 formed in the weakly 
  twisted region, it seems to be associated with strong illumination in the 
  later stage (02:22 UT and 02:24 UT on December 13) in the strongly twisted 
  region. From this result, we infer that the strong CaII illumination seems 
  to be related to the strongly twisted magnetic fields. 
 
\begin{figure}
  \epsscale{1.}
  \plotone{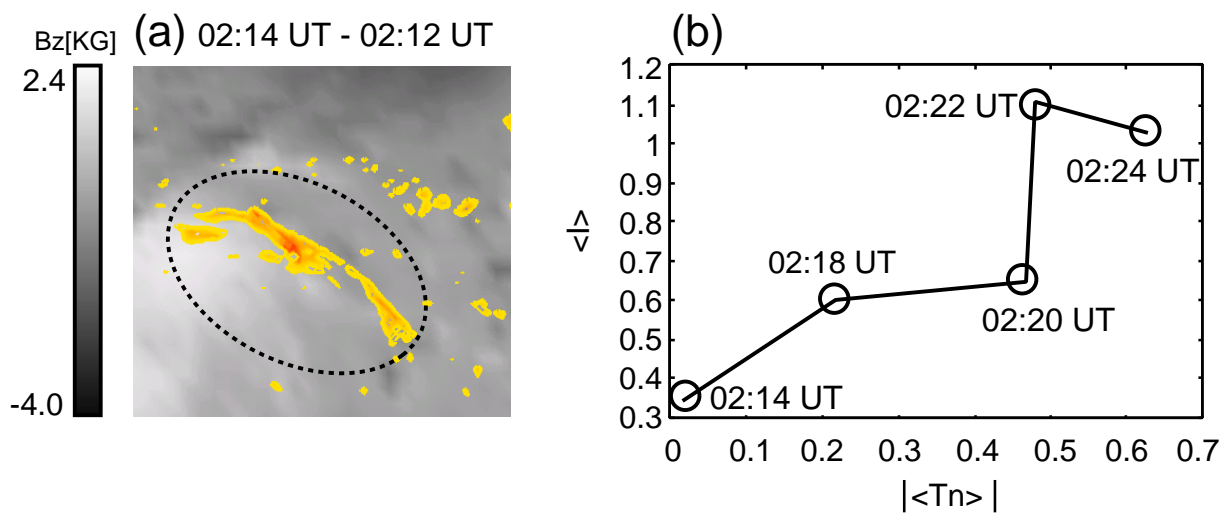}
  \caption{
           (a) Areas of strong CaII intensity ($I>$150, filled contours) 
               at 02:14 UT on December 13 plotted on a distribution map 
               of the normal component of the magnetic field (gray scale) 
               at 20:30 UT on December 12. Image covers the area 
               enclosed by the dotted square in Figure \ref{f5}(a). Northern 
               and southern ribbons cannot be distinguished at this time 
               because the two-ribbon structure had not formed yet, as 
               shown in Figure \ref{f5}. 
           (b) Profile of $\langle I \rangle$, normalized by the 
               $\langle I \rangle$ = 1000, as a function of 
               $|\langle T_n \rangle|$ at various times on December 13.
           }
  \label{f6}
  \end{figure}

  We further investigate the relationship between the CaII intensity and 
  magnetic twist value more quantitatively. We define the average 
  CaII intensity $\langle I \rangle$ and average twist value 
  $|\langle T_n \rangle|$. The CaII intensities are also based on the 
  running difference images, as shown in Figure \ref{f6}. Here ``average'' 
  indicates the average intensity in the northern ribbon, where only the 
  region of strong CaII illumination is selected by setting the threshold 
  value of the CaII intensity to $I$ = 200. However, because the two-ribbon 
  structure had not formed at 02:14 UT on December 13, we cannot distinguish 
  the northern and southern ribbons. Hence, we need to calculate these 
  average intensities using another path instead of the northern ribbon. 
  Figure \ref{f6}(a) shows the area surrounded by the dotted square in 
  Figure \ref{f5}(a); the strong CaII intensity (filled contours at 
  $I >$ 150) at 02:14 UT on December 13 is plotted on a distribution map of 
  the normal component of the magnetic field (gray scale) at 20:30 UT 
  on December 12. At this time only, $\langle I \rangle$ and 
  $|\langle T_n \rangle|$ were calculated for the strong CaII intensity 
  region surrounded by a dotted circle in Figure \ref{f6}(a). 
  Figure \ref{f6}(b) shows the profile of $\langle I \rangle$, which is 
  normalized by the $\langle I \rangle$ = 1000, as a function of 
  $|\langle T_n \rangle|$. This result clearly shows that 
  $\langle I \rangle$ suddenly increased for $|\langle T_n \rangle|$ = 0.47 
  at 02:22 UT on December 13, which indicates that a dramatic reconnection 
  occurred in the magnetic twist having values of 
  $|\langle T_n \rangle| >0.47$ because strong CaII illumination is strongly 
  correlated with magnetic reconnection (\citealt{2002A&ARv..10..313P}). 
  From these results, we find that the strong twist has a significant 
  relationship with the strong CaII intensity. 

  Note that this conclusion is not inconsistent with our previous 
  finding (\cite{2012ApJ...747..65I}), in which we showed that strong 
  X-ray intensity is not related to strongly twisted lines but rather shows 
  a good correlation with the field-aligned current. Our previous study 
  examined the core region of the sigmoid, which consisted of weakly 
  twisted field lines and was located inside the strongly twisted field 
  lines forming the elbow part of the sigmoid and revealing strong CaII 
  intensity. Therefore, the environment of the magnetic field lines making 
  up the core of the sigmoid differs from the environment we discuss in 
  this work.
   
  \subsubsection{Relaxation of the Magnetic Twist through the Flare} 
\begin{figure}
  \epsscale{1.}
  \plotone{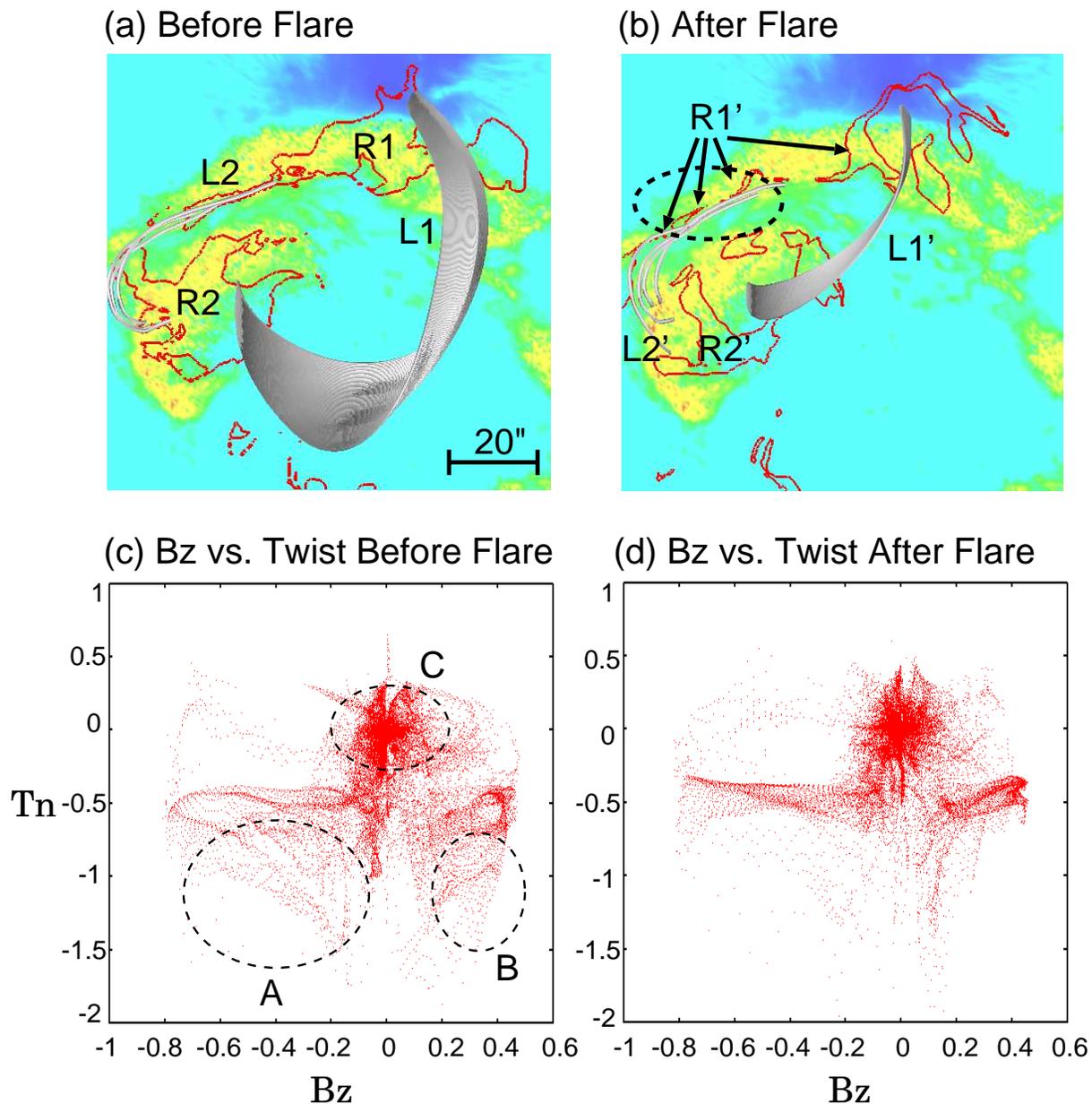}
  \caption{
           Color maps of time integration of CaII images from 02:20 UT to 
           02:40 UT on December 13. Red lines represent twist values 
           of $|T_n|$ = 0.5 from the NLFFF at (a) 20:30 UT on December 12 
           (before the flare) and (b) 04:30 UT on December 13 (after the 
           flare). R1 (R1') and R2 (R2') represent the strongly twisted 
           regions ($|T_n|>$0.5) on the negative and positive polarities, 
           respectively. L1 (L1') and L2 (L2') are selected magnetic field 
           lines connecting R1 (R1') and R2 (R2'). (c) and (d) Scatter plots 
           of twist $T_n$ (vertical axis) vs. $B_z$ (horizontal axis) before 
           and after the flare, respectively. Dashed circles A and B in (c) 
           show the regions where the strong twist was distributed before 
           the flare. Region C represents the positively twisted region of 
           approximately $|B_z| <$ 0.2 before the flare.           
          }
  \label{f7}
  \end{figure}

  We proceed with a detailed investigation of the changes in the magnetic 
  topology due to magnetic reconnection before and after the flare by 
  using the magnetic twist obtained through the NLFFF and CaII images. 
  Figures \ref{f7}(a) and (b) show the magnetic field lines (in gray) 
  before and after the flare over a time integration of the CaII images 
  from 02:20 UT to 02:40 UT on December 13, that is, from the growth to 
  the formation of the two-ribbon flare structure. The regions surrounded 
  by red lines, labeled R1 and R2 or R1' and R2', are occupied by closed 
  strongly twisted lines ($|T_n|>0.5$) from the NLFFF (a) before and (b) 
  after the flare (20:30 UT on December 12 and 04:30 UT on December 13, 
  respectively). Selected magnetic field lines, labeled L1 (L1') and 
  L2 (L2') connect the strongly twisted regions R1 (R1') and R2 (R2') 
  before (after) the flare. Note that although L1 (or L2) and L1' (or L2') 
  are plotted for the same location, there is no similarity among them.

   The strongly twisted regions R1 and R2 before the flare are clearly 
  located in the area where the CaII image showed brightening, some parts 
  of which disappeared after the flare even though regions R1' and R2'  
  remained as such. In the region where the strong twist disappeared, 
  the selected magnetic fluxes L1' and L2' seemed to relax into an 
  untwisted field relative to L1 and L2. In particular, the field line L1 
  forms a compact loop, as shown in \cite{2012ApJ...748...77S}. Because 
  this relaxation appears on the image of the strong CaII illumination, 
  magnetic reconnection may be a candidate for explaining the relaxation 
  of the magnetic twist.

   Figures \ref{f7}(c) and (d) show scatter plots of the twist $T_n$ 
  (vertical axis) vs. the normal component of the magnetic field $B_z$ 
  (horizontal axis) before and after the flare, respectively. Negatively 
  twisted lines are distributed over a wide range of the normal component 
  of the magnetic field. Dashed circles A and B indicate the strong 
  negatively twisted regions in the negative and positive polarities, 
  respectively, before the flare. Many of the points in regions A and B 
  before the flare clearly disappeared after the flare, although some    
  remained. The twisted field lines representing $T_n  \approx -0.5$ 
  seem to still be distributed over a wide range of the normal component 
  of the magnetic field after the flare. Therefore, we suggest that most 
  of the twist release is caused by field lines with at least $T_n<-0.5$. 
  On the other hand, the positively twisted regions marked by dashed circle 
  C appear for $|B_z|<0.2$ before the flare and remain as such even after 
  the flare. From this result, we clearly see again that the mixed region 
  indicating coexistence of the negatively and positively twisted lines is 
  formed at $|B_z|<0.2$, as shown in the left panels in Figure \ref{f4}.

  \section{Discussion}
   In the previous section, we discussed the buildup and release processes 
   of the magnetic twist through the huge X-class flare associated with AR 
   10930. Here, we address a possible mechanism for the onset of this flare.

   The above results (Figures \ref{f5} - \ref{f7}) showed that energy is 
   released mainly by magnetic reconnection of the field lines having a 
   strong negative twist. However, we still have not answered the question 
   regarding the flare onset mechanism that we posed at the beginning of 
   the paper. Figure \ref{f5}(a) shows that flare onset starts at around 
   02:14 UT between the positive and negative polarities of the sunspot, 
   where weakly twisted lines ($|T_n|<0.25$) appear near the PIL. Because 
   this region is also composed of a mixture of field lines having both 
   positive and negative twist, magnetic reconnection may be easily induced 
   between lines having opposite twist. The mixed-twist region was able to 
   form because positively twisted field lines were injected across the 
   pre-existing negatively twisted lines.

\begin{figure}
  \epsscale{1.}
  \plotone{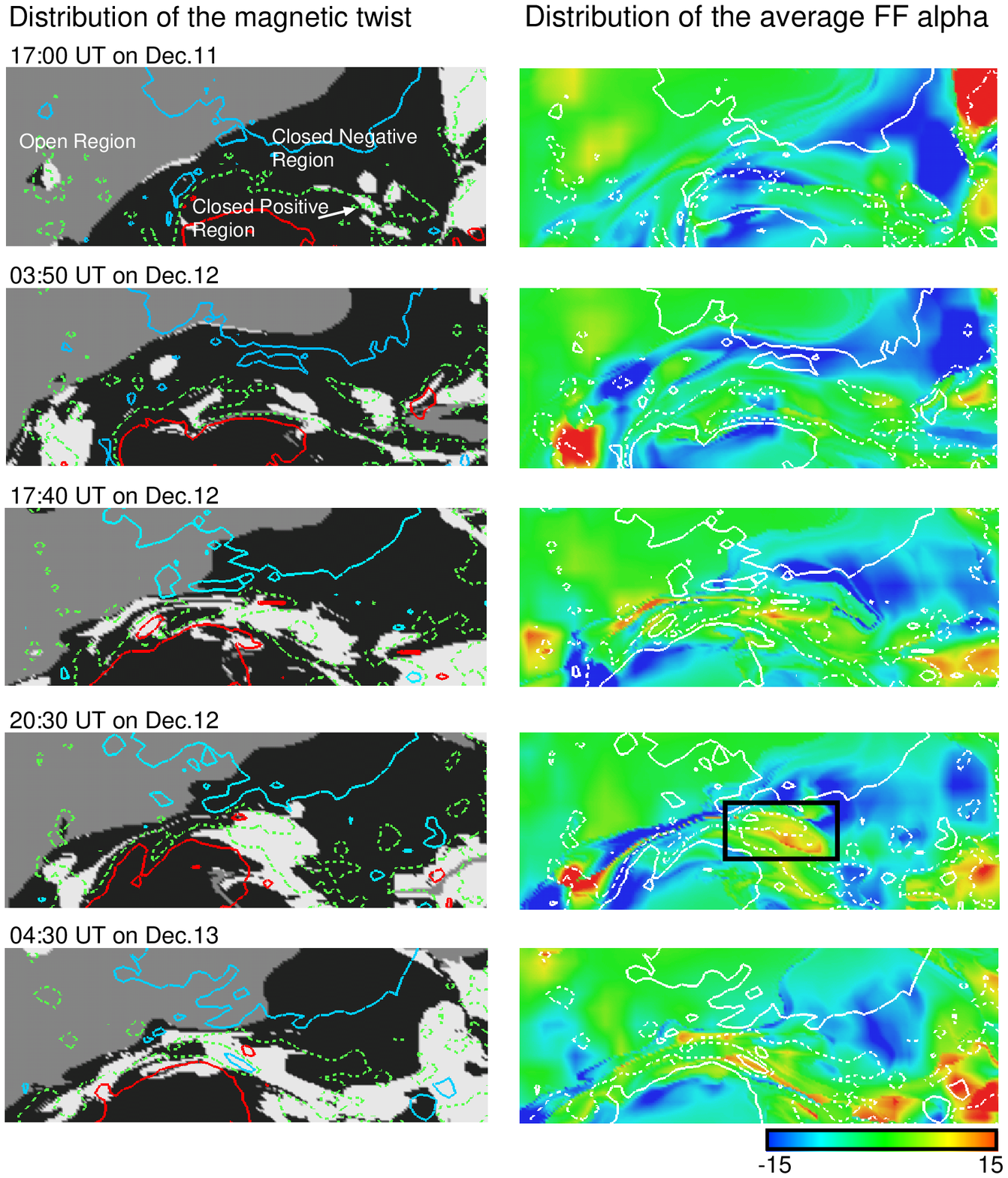}
  \caption{
               Left panels;
               Positive and negative twist distributions for closed 
               loops plotted in white and black, respectively. Gray 
               regions are occupied by open field lines. Red and blue 
               lines indicate the locations of the positive and negative 
               sunspots, respectively, which correspond to a normal 
               component of the magnetic field of 790 G and $-$790 G, 
               respectively. Dotted green lines indicate the PIL. 
               Right panels;
               Average force-free $\alpha$, denoted by $\bar{\alpha}$, 
               plotted in color. White solid lines indicate positive and 
               negative polarities; white dotted line represents the PIL, 
               which is also shown in (a).                  
                  }
  \label{f4}
  \end{figure}

\begin{figure}
  \epsscale{1.}
  \plotone{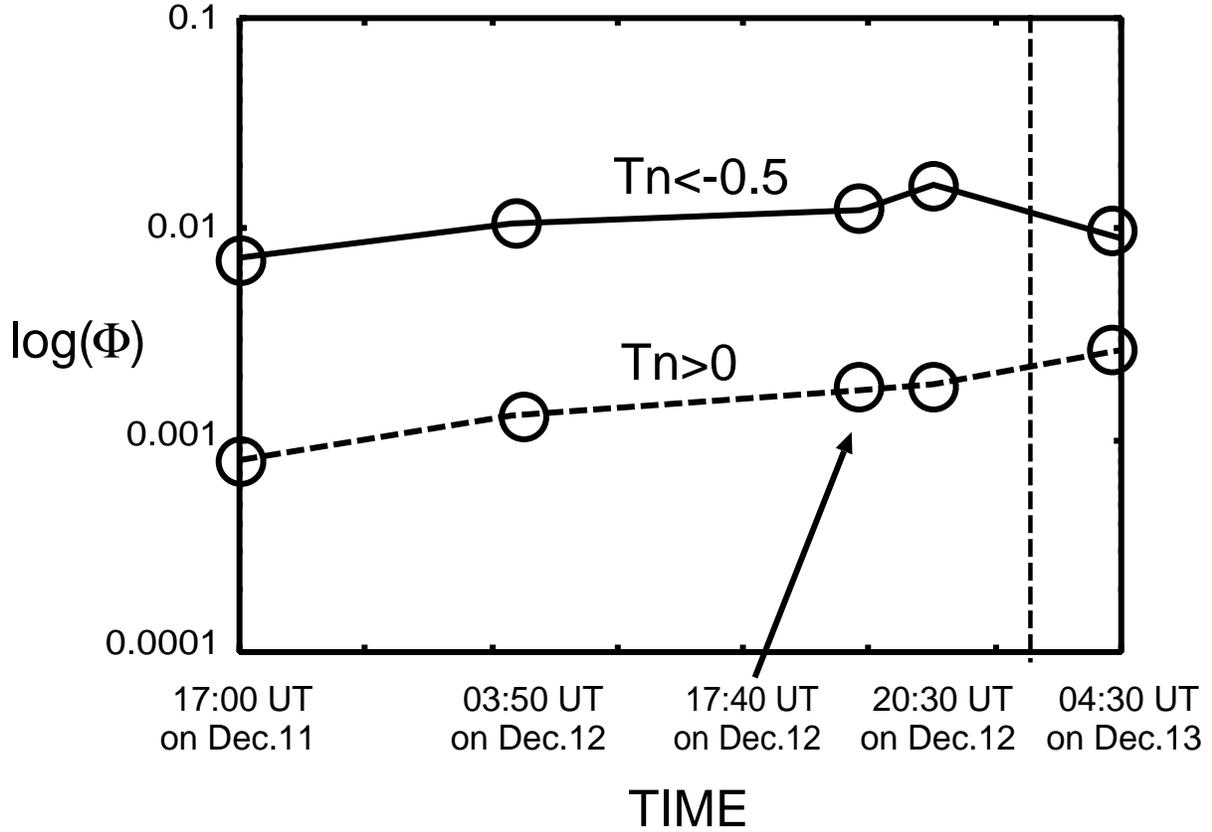}
  \caption{
           Temporal evolution of the magnetic flux, log($\Phi$) ($B_z>0$),  
           for $T_n>0$ (dashed line) and $T_n<-0.5$ (solid line), 
           respectively, corresponding to the areas displayed in 
           Figures \ref{f2} and \ref{f4}, respectively. These values are 
           normalized by $10^{24}$ (Mx) (=$\int B_0 dS$; $dS$ represents 
           an surface element in the bottom surface). Vertical dashed 
           line indicates the time at which the flare occurred.          
          }
  \label{f8}
  \end{figure}

   Further, we investigate the temporal evolution of the magnetic flux 
   corresponding to the positive twist and compare it with that of the  
   magnetic flux corresponding to the negative twist. Figure \ref{f8} 
   shows the temporal evolution of the magnetic flux, 
   log ($\Phi$) ($B_z >0$), for $T_n < -0.5$ and $T_n > 0$, which are 
   represented by solid and dashed lines, respectively, corresponding 
   to the areas displayed in Figures \ref{f2} and \ref{f4}, respectively. 
   The vertical dashed line indicates the flare onset time. We found 
   that most of the magnetic twist producing the flare was greater than 
   a half-turn in the negative sense ($T_n < -0.5$). The time profiles 
   of both magnetic fluxes show an increase as the flare onset time 
   approached. However, the magnetic flux related to $T_n<-0.5$ decreased 
   dramatically around the flare onset time. We believe that this decrement 
   in the magnetic flux related to the negative twist is due to the 
   relaxation via magnetic reconnection shown in Figures \ref{f5}-\ref{f7}. 
   In contrast, the magnetic flux corresponding to $T_n>0$ increased 
   continuously before and after flare onset.   
   
   In previous studies, \cite{2011ApJ...738..161I} and 
   \cite{2012ApJ...747..65I} indicated that the magnetic configuration 
   occupying the strongly twisted regions before the flare was robust 
   against kink mode instability; therefore, this AR cannot be destabilized 
   through kink instability. Hence, flare onset is triggered via another 
   mechanism: breaking of the equilibrium condition of the strongly 
   twisted field; the buildup of positive twist may be a candidate cause 
   of that collapse. Consequently, we can say that the increase in the 
   magnetic flux due to injection of twist of the opposite sign across 
   the pre-existing field is an important process for understanding the 
   flare trigger mechanism.
   
   We again look at the distribution maps of the twisted and average 
   force-free $\alpha$ (denoted as $\bar{\alpha}$) at 20:30 UT on December 12 
   in the left and right panels in Figures \ref{f4}. Although the twist value 
   in the mixed-sign region marked by the black square in Figure \ref{f1}(a) 
   is weak relative to the negative twist values distributed at a considerable
   distance from the PIL (Figure \ref{f1}(b)), the strong $\bar{\alpha}$ is 
   injected into the region occupied by this weak mixed-sign field (for 
   example, the region surrounded by the solid square in the right panels in 
   Figure \ref{f4}). The twist value is small in this region because it 
   depends strongly on the field line length, even though strong force-free 
   $\alpha$ exists at their footpoints. \cite{1997ApJ...481..973P} reported a 
   strong correlation between the force-free $\alpha$ and the geometric shear 
   angle $\theta$ (equation (5.3.18) in \citealt{2004psci.book.....A}) from 
   the soft X-ray loops observed by the Soft X-Ray Telescope (SXT) on 
   {\it Yohkoh}. Therefore, this result implies that the buildup of strong 
   magnetic shear of opposite sign may be related to flare onset, even though 
   the field lines are extremely short (see  Figure \ref{f1}(c)).  However, we
   cannot identify the essential process related to the amount of positively 
   twisted magnetic flux or the strength of the average force-free $\alpha$. 
   We need further statistical analysis using data with a higher time cadence 
   or numerical experiments to answer this question.
   
   Some authors have already proposed theoretical flare/CME 
   models in which small-scale reconnection in the local area at 
   the lower corona leads to large-scale reconnection, releasing 
   the magnetic energy. In a 2D model, \cite{2000ApJ...545..524C} 
   and \cite{2005ApJ...634..663S} successfully interpreted an 
   eruption as magnetic reconnection between the emerging flux 
   and the pre-existing magnetic flux, which leads to a loss of 
   equilibrium in the flux tube; subsequently, another magnetic 
   reconnection is induced in the pre-existing lines under the 
   eruptive flux rope.
   
   On the other hand, \cite{2004ApJ...610..537K} conducted a 3D 
   simulation to determine the role of local instability to determine 
   whether it leads to an ejective eruption. They reported that tearing 
   instability occurs at the lower corona where the reversed magnetic 
   shear field is formed against the overlying field and eventually 
   leads to ejective eruption through a nonlinear feedback process, 
   that is the mutual interaction between large-scale strongly twisted 
   lines and a small-scale region of mixed twist 
   (\citealt{2004ApJ...610..537K} ). They also proposed that the key 
   process of flare onset is magnetic reconnection between field lines 
   having different twist in the local area of the lower corona, which 
   becomes an important cause of large-scale flares. Thus, our result 
   shows behavior similar to that of the theoretical model proposed 
   by \cite{2004ApJ...610..537K}. However, it is much difficult to 
   conclude an answer of the onset mechanism in term of the only 
   quasi-static pictures obtained from NLFFF.    
     
   Although we suggest that the formation of two different regions is 
   important for flare onset, the detailed formation process is still 
   not well known. Nevertheless, 
   \cite{2010ApJ...720.1102P},
   \cite{2011ApJ...740...19R}, and
   \cite{2011ApJ...743...33R} recently suggested that
   the process of building up positively twisted lines over pre-existing 
   negatively twisted lines seems to be a possible component of the 
   flare onset mechanism. \cite{2008ApJ...687..658W} and 
   \cite{2010ApJ...719..403L} indicated that flux emergence is a key 
   process in the formation of these structures and is also responsible 
   for flare onset. The key problems are how the mixed-helicity region 
   is formed on the local area near the PIL and how it plays a key role 
   in exciting a flare. A flux emergence simulation may address these 
   questions in the future.

  \section{Summary}
   In this paper, we investigated the temporal evolution of the 3D magnetic 
   structure of AR 10930. We examined the buildup and release processes of 
   the magnetic twist associated with this AR using a time series of the 
   vector fields before and after a flare. Our analysis is based on the 
   NLFFF extrapolation method, which is applied to time series data obtained 
   from SP/{\it Hinode}. We also suggested a possible flare onset mechanism 
   in terms of magnetic reconnection between field lines having opposite twist.
   
   We found that the 3D magnetic structure of this AR can be roughly 
   separated into two regions containing field lines having different 
   twist. One is occupied by field lines of strong negative twist, 
   ({\it i.e.}, left-handed twist) rooted in the regions of both polarities 
   and is located at a considerable distance from the PIL. The other is of 
   mixed polarity and contains both positively and negatively twisted lines 
   near the PIL between two sunspots. We investigated the temporal  
   evolution of the magnetic twist ({\it i.e.}, from its buildup to its 
   release) associated with this AR using time series NLFFF. We found that 
   a day before flare onset, most regions were occupied by negatively 
   twisted field lines. Positively twisted lines were also built up near 
   the PIL within one day; consequently, the mixed region was formed, 
   which was eventually occupied and dominated by positively twisted lines 
   after the flare.
  
   For flare onset, we suggest the importance of the buildup near the 
   PIL of twisted magnetic lines of different sign compared to the 
   ambient field. In this situation, magnetic reconnection can be induced 
   between field lines having different twist, which would destroy the 
   equilibrium of the magnetic configuration. We also found that the main 
   relaxation process in the flare dynamics is caused by magnetic 
   reconnection in strongly negatively twisted lines by comparing CaII 
   images with the NLFFF. This scenario is similar to a previous model 
   proposed by \cite{2004ApJ...610..537K}. Unfortunately, we cannot 
   extract much more information from the NLFFF and conclude a clear 
   answer for the onset mechanism because it is useful for magnetic 
   structures in a quasi-static state. Thus, we still have problems such 
   as the transition process from a stable condition to an unstable or 
   non-equilibrium condition and the 3D dynamics of magnetic reconnection 
   in the solar corona. We believe that MHD simulations may provide strong 
   clues regarding these points.
   
   In addition, this study is essentially limited by the low time cadence 
   of the SP data ($\sim$10 h) used in this analysis. To achieve a more 
   comprehensive view of the temporal evolution of the 3D magnetic structure, 
   high-cadence data are essential. A new solar physics satellite, 
   {\it Solar Dynamics Observatory} ({\it SDO}) was launched recently by 
   the National Aeronautics and Space Administration (NASA). The Helioseismic 
   and Magnetic Imager (HMI) on board {\it SDO} can provide vector 
   magnetograms with higher time cadence and a larger field of view compared 
   to the data obtained from {\it Hinode}. The detailed time evolution of the 
   magnetic field lines and accurate reproduction of the location of the 
   separatrix are important issues for the understanding of the flare onset 
   and dynamics. We will deepen our understanding of the onset of solar 
   flares by combining high temporal resolution data from {\it SDO} with 
   high spatial resolution data obtained by {\it Hinode}.

  \acknowledgments 
  S.\ I.\ is grateful to Prof.\ K.\ Kusano for the many constructive comments 
  and useful discussions. Authors acknowledge the anonymous referee for 
  his/her careful review and constructive comments. S.\ I.\ was supported 
  by the International Scholarship of Kyung Hee University. This study was 
  supported by the WCU (World Class University) program (R31-10016) and by 
  Korea Meteorological Administration through National Meteorological 
  Satellite Center as well as by the Basic Science Research Program 
  (2010-0009258, PI: T. Magara) through the National Research Foundation of 
  Korea. G.\ S.\ C.\ was supported by the Korea Research Foundation grant 
  funded by the Korean Government (KRF-2007-313-C00324). D.S. is supported 
  in part by the Special Postdoctoral Researchers Program at RIKEN. The 
  computing, data analysis and visualization were performed using the 
  OneSpaceNet in the NICT Science Cloud. Hinode is a Japanese mission 
  developed and launched by ISAS/JAXA, with NAOJ as domestic partner and NASA 
  and STFC (UK) as international partners. It is operated by these agencies 
  in co-operation with ESA and NSC (Norway). We thank the MDI consortia for 
  providing data. SoHO is a mission of international cooperation between ESA 
  and NASA. The ambiguity resolution code used here was developed by 
  K.\ D.\ Leka, G.\ Barnes, A.\ Crouch with NWRA support from SAO under NASA 
  NNM07AB07C.

\clearpage



  \clearpage

  \clearpage

  \clearpage
  
  \clearpage

  \clearpage

  \clearpage

  \clearpage

  \end{document}